\documentclass[twocolumn, superscriptaddress]{revtex4-1}
\usepackage{bm}
\usepackage[pdftex]{graphicx}
\usepackage{color}
\usepackage{soul}
\usepackage{xcolor}
\usepackage{amsmath}
\usepackage{amssymb}
\usepackage{url}
\usepackage[utf8]{inputenc}

\begin{document}

\title{Extended correlations in the critical superheated solid}

\author{Vivianne Olgu\'in-Arias}
\email{vivianne.olguin.a@gmail.com}
\affiliation{Departamento de F\'isica, Facultad de Ciencias Exactas, Universidad Andres Bello. Sazi\'e 2212, piso 7, Santiago, 8370136, Chile.}

\author{Sergio Davis}
\affiliation{Comisión Chilena de Energía Nuclear, Casilla 188-D, Santiago, Chile}
\affiliation{Departamento de F\'isica, Facultad de Ciencias Exactas, Universidad Andres Bello. Sazi\'e 2212, piso 7, Santiago, 8370136, Chile.}

\author{Gonzalo Gutiérrez}
\affiliation{Grupo de Nanomateriales, Departamento de F\'{i}sica, Facultad de Ciencias, Universidad de Chile, Casilla 653, Santiago, Chile}

\begin{abstract}
Metastable states in first-order phase transitions reveal interesting behavior about a wide range of systems in statistical mechanics, including spin systems, cellular automata and
condensed matter systems. These metastable states are often observed in a microcanonical setting, where they manifest long-range correlations due to collective effects. In this work
we show the existence of long-range potential energy correlations between atoms in a microcanonical superheated Lennard-Jones crystal prior to homogeneous melting. Our results suggest that the cooperative motion made possible by the presence of vacancy-interstitial pairs above the melting temperature induces effective long-range interatomic forces even beyond the fourth neighboring layer.
\end{abstract}

\maketitle

\section{Introduction}
\label{introduction}

Melting is a common phenomenon in our daily life, and a clear example of a first-order, or discontinuous, phase transition. Although it is
understood in thermodynamic (macroscopic) terms, the transition itself has eluded a complete description from the point of view of
microscopic dynamics. In recent decades, computer simulation methods such as molecular dynamics (MD) and Monte Carlo methods,
together with modern statistical techniques, have become important theoretical tools to address the challenge of understanding the melting
mechanism from an atomistic point of view~\cite{Jin2001,Luo2003,Forsblom2005}.

The melting process in most cases turns out to be an heterogeneous process, since it is commonly triggered on the surface of the crystal. Melting under heterogeneous conditions occurs exactly at the melting temperature $T_m$ and is well described by the classical nucleation theory~\cite{Bai2005b,Luo2007}.
However, under ideal conditions, i.e. without surfaces or defects, it is possible to carry out this process homogeneously, in such a way that a superheated, metastable solid state can be reached, far above $T_m$.

Eventually, as we increase the energy of the crystal, we reach a critical temperature $T_{LS}$, called the limit of superheating, such that
the solid structure cannot exist at a higher temperature than this without being spontaneously transformed into liquid. This process of
collapse of the crystalline state is believed to originate from the diffusion of mobile vacancy-interstitial pairs that occurs as the solid
surpasses $T_m$. In the superheated solid state, the atoms in the crystal can temporarily occupy interstitial sites, creating
vacancies~\cite{Belonoshko2007} and allowing neighboring atoms to diffuse occupying the vacant sites~\cite{Bai2008, Davis2011, Gallington2010}.
Beyond $T_{LS}$, the solid is no longer in the metastable state but reaches an unstable state, that may be referred to as the \emph{critical
superheated solid}.

In stark contrast with the study of metastable states in classical spin Hamiltonians~\cite{Binder1974,Tomita1992,Moreno2018}, cellular automata~\cite{Wu2008}, glassy systems~\cite{Monasson1995} and other models~\cite{Pluchino2004}, the statistical mechanical description of the microcanonical superheated solid state is clearly lacking. The case
of the supercooled liquid, which is a glassy metastable state, is surprisingly similar to the superheated solid state. There, long-range correlations has been observed, both experimentally
and via computer simulation~\cite{Doliwa2000, Bakai2002, Bakai2004, Weeks2007, Szamel2011, Flenner2015b}, associated with cooperative motion of the atoms. Therefore, a natural
question is the existence of similar behavior in the superheated solid state.

Traditionally, correlation is measured through the use of the Pearson correlation metric. This, however, is only able to detect certain types of dependencies between variables. On
the other hand, the \emph{mutual information} metric~\cite{CoverThomas2006} measures the amount of information that one variable contains about to the other, where through its use it
is possible to establish the existence of correlation between the two. The mutual information metric has been previously used in studies of phase transitions and metastability in several systems, in particular spin systems and glasses~\cite{Wicks2007, Wilms2011, Wilms2012, Dunleavy2015, Alba2016}.

A particularly powerful approach to study the superheated solid state in MD simulation is the so called Z-method~\cite{Belonoshko2006}.
Originally proposed as a practical method to compute melting points in atomistic simulation, it has also become a way to explore the behavior
of the superheated solid state and explain the physical meaning of the maximum superheating temperature $T_{LS}$. In this method, the
perfect lattice configuration is used as the initial state, and the system is simulated entirely in the microcanonical (NVE) ensemble.

In this work, the behavior of a metastable Lennard-Jones crystal was studied from the observation of atoms and their neighbors immediately before the melting process using molecular dynamics simulations. As stipulated in the Z-method, the calculations were carried out under microcanonical conditions, that is, at fixed energy, volume and number of particles.
By means of the atomistic study of superheating in the LJ system, we have managed to contribute to the understanding of the
metastable solid phase, presenting evidence of correlations at extended distances that we have measured using the mutual information metric.

This paper continues as follows. In Sec. \ref{computational}, the computational procedure is described in detail. Sec. \ref{results} presents
the main simulation and statistical results, while the paper concludes in Sec. \ref{concluding} with some final remarks.

\section{Computational Procedure}
\label{computational}

In this work we performed a number of microcanonical (NVE) MD simulations at different initial temperatures $T_0$, equivalent
to different total energies of the system through the relation $$E = \frac{3}{2}Nk_B T_0 + \Phi_0,$$ with $\Phi_0$ the potential energy of the ideal crystal. In our case, $\Phi_0$= 556.74 eV. We considered as a starting point for these MD simulations a perfect crystal of argon, corresponding to a face-centered-cubic (FCC) structure, and interacting through a standard Lennard-Jones (LJ) potential,

\begin{align}
\Phi(\bm{r}_1, \ldots, \bm{r}_N) & = \sum_i \Phi_i(\bm{r}_1, \ldots, \bm{r}_N) \nonumber \\
                                 & = \sum_i\left[\frac{1}{2}\sum_{j\neq i}\varphi(|\bm{r}_i-\bm{r}_j|)\right],
\end{align}
where $\varphi(r)$ is the pair potential function,
$$\varphi(r) = 4\epsilon\left[\Big(\frac{\sigma}{r}\Big)^{12}-\Big(\frac{\sigma}{r}\Big)^6\right].$$

The crystal is simulated at high density in order to enhance the superheating effect, using a lattice constant $a=4.2$ \AA~and the usual LJ parameters for argon, namely $\epsilon/k_B=119.8$ K and $\sigma=3.41$ \AA. All MD simulations were performed using the LPMD package~\cite{Davis2010}, with a timestep $\Delta t=$1 fs. For every value of initial temperature $T_0$ we used a total simulation time of 80 ps, corresponding to 80000 steps.

\begin{figure}[h!]
\begin{center}
\includegraphics[width=0.95\columnwidth]{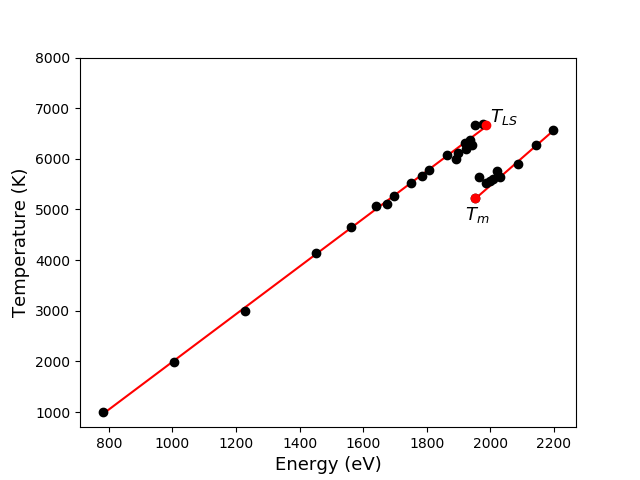}
\caption{Isochoric (Z) curve obtained from 31 different 80 ps simulations of high-density Ar, with initial temperatures $T_0$ ranging from 2000 to 12600 K. From this Z-curve
we can determine $T_m \sim$ 5231 K and $T_{LS} \sim$ 6695 K. }
\label{fig-Zmeth}
\end{center}
\end{figure}

As shown in Fig.\ref{fig-Zmeth}, through the use of the Z-method we have obtained an isochoric curve with a melting temperature $T_m=5231$ K and superheating temperature $T_{LS}=6695$ K. Also a critical energy $E_{LS}=1976$ eV was obtained, such that it is impossible to administer additional energy without the system melting spontaneously.

The regions of thermodynamic stability are given by $E <$ 1600 eV ($T < T_m$), where the solid phase is stable, and
$E >$ 1976 eV where the liquid phase becomes stable. For energies between 1600 eV and 1850 eV the system is found in a
metastable, superheated solid phase, while in the region between $E$=1850 eV and $E$=2050 eV, the system exists in either of two phases,
namely superheated solid or liquid.

In the next section we present the results obtained from MD simulation, which we have used to establish the probability distribution
describing the potential energy per atom $\phi$ as a function of total energy $E$. Moreover, we present evidence for the existence of
increasing correlation between atoms at long-range distances as the temperature increases, as measured through the mutual information metric.

\section{Results}
\label{results}

\begin{figure}[h!]
\begin{center}
\includegraphics[width=0.98\columnwidth]{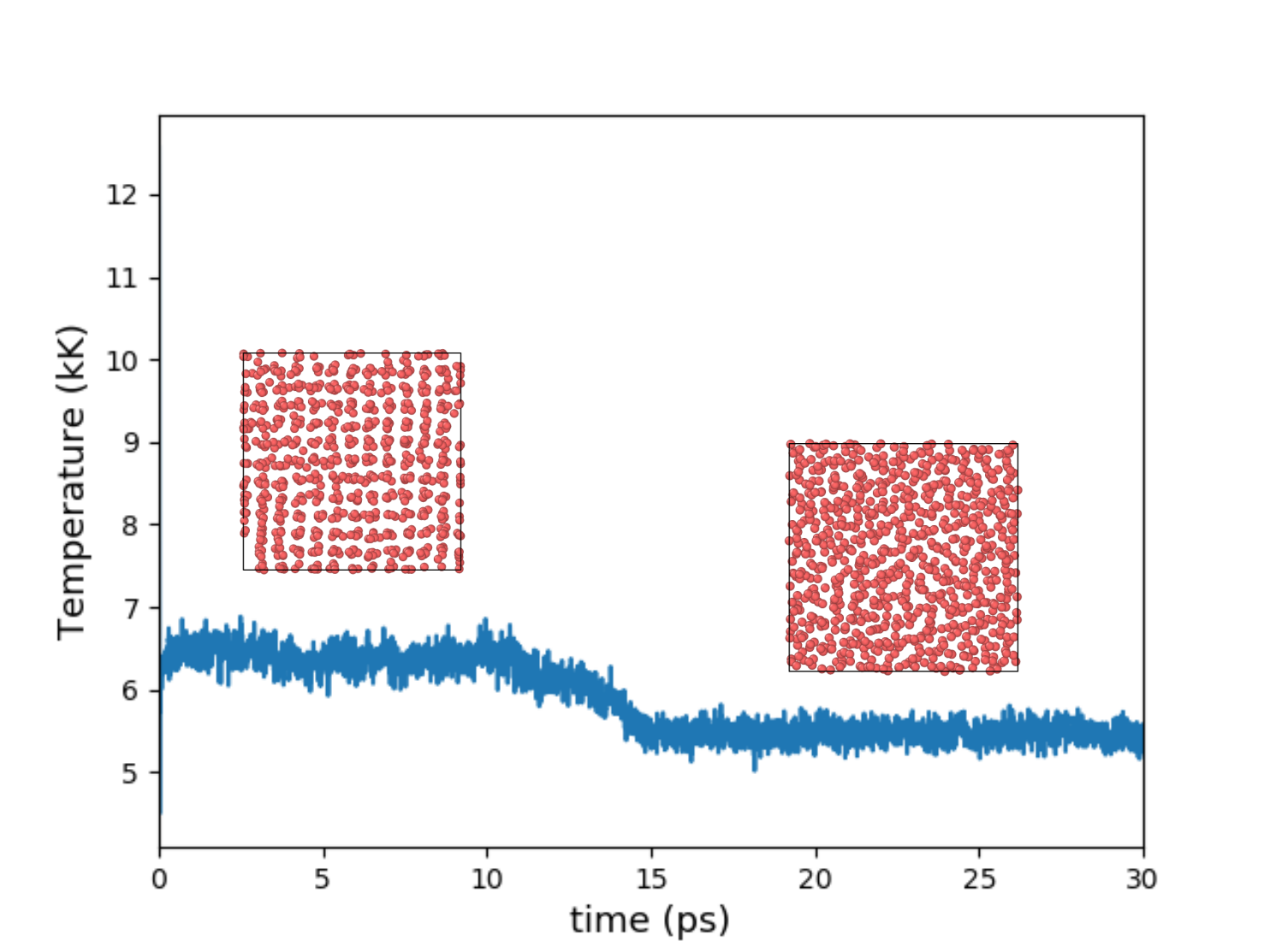}
\caption{Instantaneous temperature $T(t)$ for a simulation at an initial temperature $T_0=12600$ K. Only the first 30 ps of the simulation are shown, as after the melting process occurs at time $t_w$=9 ps the system remains in the liquid phase.}
\label{fig-melt}
\end{center}
\end{figure}

Fig. \ref{fig-melt} shows a typical microcanonical simulation during which melting is triggered spontaneously. The instantaneous temperature of the system, originally prepared at $T_0=12600$ K from an ideal crystalline structure, first drops to approximately 6500 K due to the equilibration of potential and kinetic energies. At this point the system remains in a metastable, superheated solid state. After $t_w$=9 ps for this particular simulation, the crystal suddenly collapses and the temperature drops to about 5500 K. For several simulations at the same total energy we obtained random melting times ranging from 1 to 80 ps, which is consistent with previous studies of the kinetics of homogeneous melting~\cite{Alfe2011,Davis2018d}. In terms of the atomistic dynamics, the picture is as follows. As the system begins to evolve from the ideal crystalline structure, the atoms begin to move away from their stable local potential energy minima. For temperatures above $T_m$ the energy per atom is such that thermally-activated point defects start to form~\cite{Balluffi2005}. This increase of the number of defects and their mobility results in a cooperative motion involving several atoms.

\begin{figure}[h!]
\begin{center}
\includegraphics[width=0.98\columnwidth]{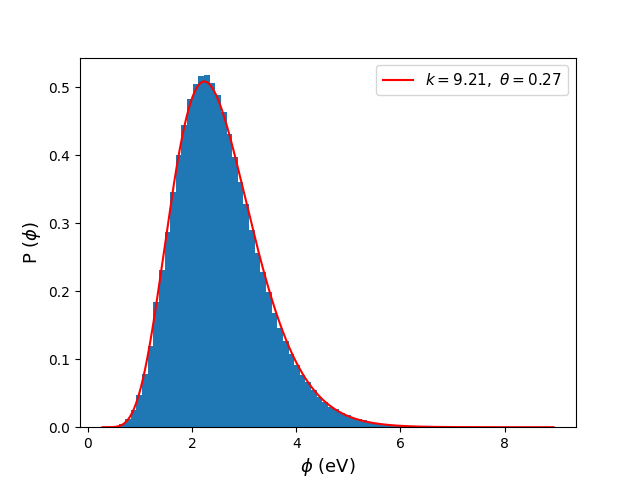}
\caption{Histogram of potential energy per atom for a system at $T_0=10000$ K (blue filled region) and gamma statistical distribution with parameters $k$=9.21 and $\theta$=0.27 eV (red curve).}
\label{fig_dist1}
\end{center}
\end{figure}

In order to reveal the existence of extreme events and interatomic correlations in the superheated phase, we have studied the
statistics of the potential energy per atom $\phi$. In the low-temperature, harmonic approximation of a solid, each atom moves independently
as an harmonic oscillator anchored to its equilibrium position. It can be shown that for the microcanonical ensemble in the limit $N \rightarrow \infty$ (see the Appendix
for more details), the probability distribution function of $\phi$ for non-interacting particles is given by
\begin{equation}
P(\phi_i | E) = \frac{1}{Z_1(\beta)}\exp(-\beta \phi_i)\Omega_1(\phi_i),
\label{eq_phi_boltz}
\end{equation}
with $\beta=3N/2E$. In the superheated solid, however, we have found that $\phi$ follows a gamma distribution,
\begin{equation}
P(\phi|E) = P(\phi|k, \theta) = \frac{\exp(-\phi/\theta)\phi^{k-1}}{\Gamma(k)\theta^k},
\label{eq_gammadist}
\end{equation}
with energy-dependent parameters $k=k(E)$ and $\theta=\theta(E)$. Fig. \ref{fig_dist1} shows an example of this distribution
for $T_0$=10000 K, having a most probable energy given by the mode of the gamma distribution, $\phi^*=(k-1)\theta$, in this case
corresponding to $\phi^*$=2.217 eV.

\begin{figure}[h!]
\begin{center}
\includegraphics[width=0.95\columnwidth]{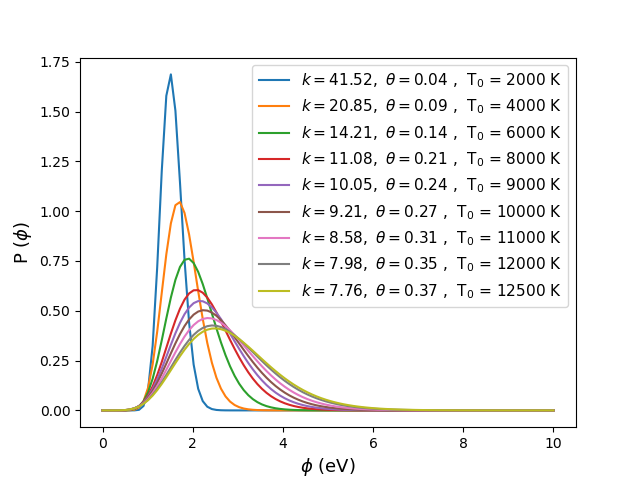}
\caption{Probability density function of single-atom potential energies $P(\phi|E)$ for different initial temperatures $T_0$ between 2000 K and 12500 K. The nearly Gaussian distribution for $T_0=$2000 K changes to a long-tailed distribution indicating correlations as the temperature $T_0$ increases.}
\label{fig_dist2}
\end{center}
\end{figure}

Fig. \ref{fig_dist2} shows the probability density function $P(\phi|E)$ for different initial temperatures ranging from
$T_0=$2000 K up to $T_0=$12500 K. Here we see that the observed $P(\phi|E)$ is incompatible with Eq. \ref{eq_phi_boltz} for independent
oscillators, because $k$ depends on the total energy, and it is in fact described by
\begin{equation}
P(\phi_1|E) = \frac{1}{Z(\beta)}\exp(-\beta \phi_1)\omega_1(\phi_1; E)
\label{eq_phi_omega1}
\end{equation}
as shown in the Appendix. Therefore, we can conclude at this point that the potential energies of different atoms are correlated up to some distance to be determined.

While at low temperatures the potential energy values are in fact distributed following a Gaussian distribution, clearly indicating independent
behavior compatible with Eq. \ref{eq_phi_boltz}, as the temperature of the system increases the atoms in the lattice break their symmetry. In this phase the solid is in
superheated phase, producing a correlation of potential energies between nearest neighbors and even larger distances; these results are consistent with Eqs. \ref{eq_phi_omega1}
and \ref{eq_gammadist}. Note that in no case we observe a bimodal distribution of potential energies, as might be expected in the case of nucleation of a liquid
phase inside a solid matrix.

\begin{figure}[h!]
\begin{center}
\includegraphics[width=0.95\columnwidth]{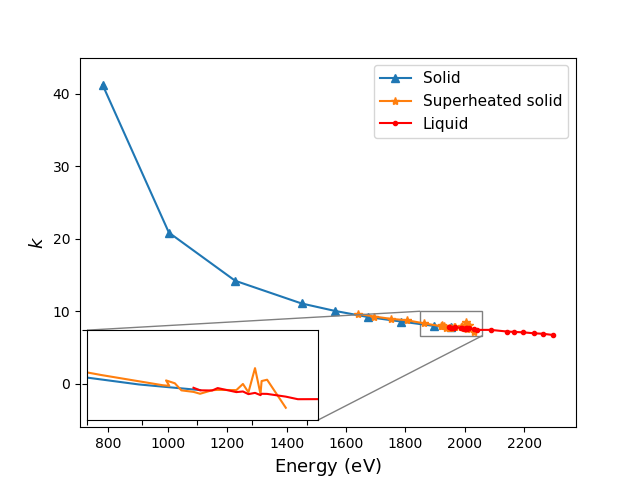}
\caption{Shape parameter $k$ as a function of the energy $E$. The solid, light blue line represents the system in its solid phase, the yellow line in its superheated solid phase and the red line in liquid phase. From the energy $E$=1947 eV to $E$=2031 eV, the system is in the critical superheated phase.}
\label{fig_dist3}
\end{center}
\end{figure}

\begin{figure}[h!]
\begin{center}
\includegraphics[width=0.95\columnwidth]{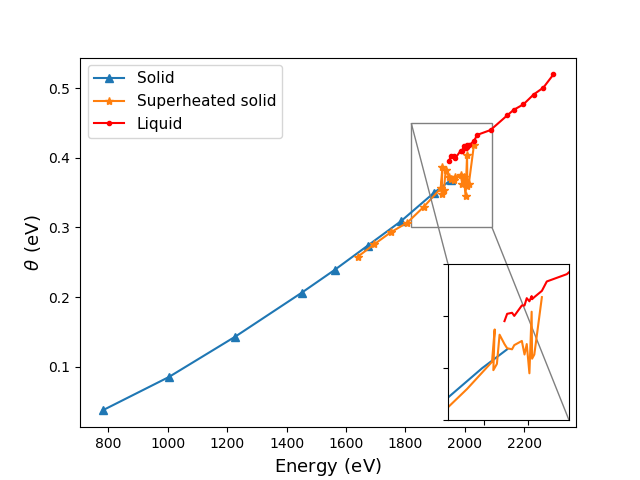}
\caption{Scale parameter $\theta$ as a function of the energy $E$. The solid, light blue line represents the system in its solid phase, the yellow line in its superheated solid phase and the red line in liquid phase. From the energy $E$=1947 eV to $E$=2031 eV the system is in the critical superheated phase.}
\label{fig_theta}
\end{center}
\end{figure}

\begin{figure}[h!]
\begin{center}
\includegraphics[width=0.95\columnwidth]{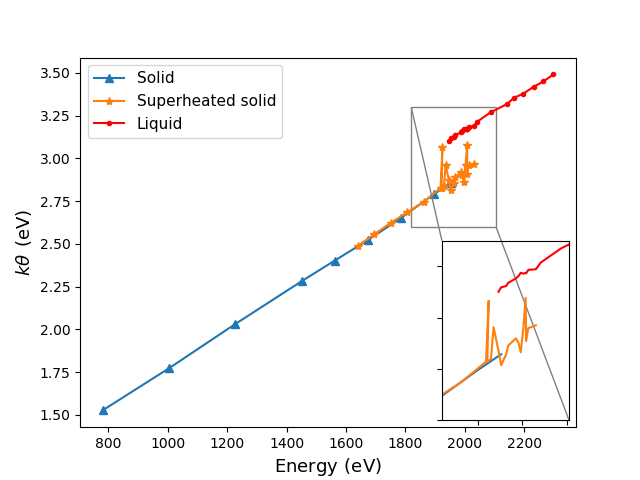}
\caption{Expected value $\big<\phi\big>_E=k\theta$ as a function of energy $E$. The solid, light blue line represents the system in its solid phase, the yellow line in its superheated solid phase and the red line in liquid phase. From the energy $E$=1947 eV to $E$=2031 eV the system is in critical superheated phase.}
\label{fig_ktheta}
\end{center}
\end{figure}

In Figs. \ref{fig_dist3}, \ref{fig_theta} and \ref{fig_ktheta} an energy of superheating between 1640 and 1936 eV can be observed. It is important to notice that the parameters
$k$ and $\theta$ present a dependence on the total energy of the system. From a total energy $E=$1947 eV upwards, the system is present in one of two possible phases: critical superheated solid and liquid. Here we can see two branches where the parameters $k$ and $\theta$ are found in the same phase of the superheated solid. The variation of these parameters with the total energy is an important indicator of the existence of correlations between atoms, where possibly every atom depends on the behavior of its nearest neighbors and even beyond third-nearest neighbors.

Fig. \ref{fig_dist3} shows the behavior of the shape parameter $k$ as a function of the total energy $E$. As the system does not behave
like a collection of independent harmonic oscillators, but is correlated, the \emph{effective} degrees of freedom are now determined by
the ``density of states'' factor in Eq. \ref{eq_gammadist}, governed by the exponent $k(E)$. As seen in the figure, for higher energies
$k$ decreases, hence a reduced number of configurations are available for each atom at a given potential energy $\phi$ which
again is suggestive of a cooperative motion.

On the other hand, Fig. \ref{fig_theta} shows that the scale parameter $\theta$ does not present a continuous behavior as a function of energy.
The discontinuity is essentially due to the latent heat per atom, which corresponds to the change in potential energy as the system undergoes
the phase transition.

Fig. \ref{fig_ktheta} shows the expected value of potential energy per atom as a function of the total energy $E$, that is,
$\big<\phi\big>_E$, which is given by $k\theta$ in terms of the Gamma parameters. We can clearly see that $\big<\phi\big>_E$
increases linearly with $E$ in both the solid and liquid branches (outside the metastable region), as expected in a system
with constant specific heat in their stable phases.

\begin{figure}[h!]
\begin{center}
\includegraphics[width=0.95\columnwidth]{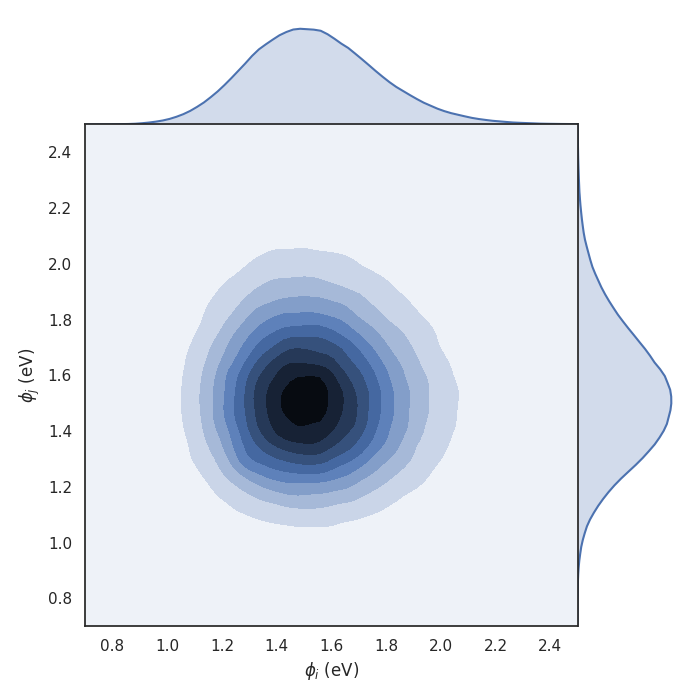}
\caption{Kernel density estimation of the joint distribution $P(\phi_1,\phi_2|E)$ for a distance $r$ = 7.25 \AA~and initial temperature $T_0=$2000 K. At low temperature the system does not present a correlation, distributing the potential energies by means of a Gaussian distribution in 2D.}
\label{fig_den2000K}
\end{center}
\end{figure}

\begin{figure}[h!]
\begin{center}
\includegraphics[width=0.95\columnwidth]{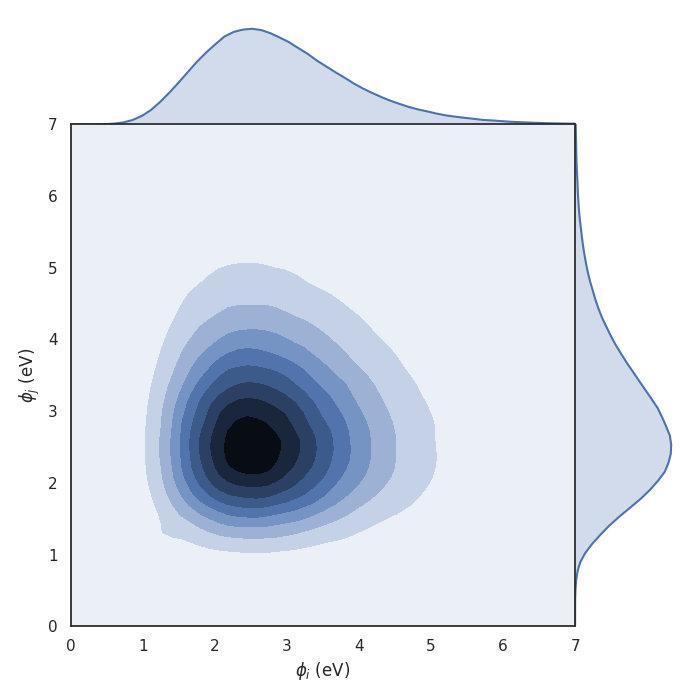}
\caption{Kernel density estimation of the joint distribution $P(\phi_1,\phi_2|E)$ for a distance $r$ = 7.25 \AA~and initial temperature $T_0=$12500 K. At high temperatures the system presents a correlation the potential energies are distributed according to a Gamma distribution.}
\label{fig_den12500K}
\end{center}
\end{figure}

From Eq. \ref{eq_Pphi12} for $P(\phi_1,\phi_2|E)$ we performed a non-parametric estimation of the joint distribution using
the kernel density estimation (KDE) method~\cite{BailerJones2017}, considering data from the MD simulations where a set of
potential energies $\phi_1, ..., \phi_N$ are divided into pairs at given distances $r$. This kernel density estimation is
shown in Figs. \ref{fig_den2000K} and \ref{fig_den12500K} for a dataset of potential energies at initial temperatures $T_0$=2000 K
and $T_0$=12500 K, respectively at a distance of $r$ = 7.25 \AA. For $T_0=$2000 K, the potential energy pairs follow a
bivariate Gaussian~\cite{BailerJones2017} with almost null correlation. However, for $T_0=$12500 K a kind of directionality in the
shape of the bidimensional distribution can be seen, in a consistent manner with the marginal Gamma distributions, which confirms the presence
of correlations between distant pairs of atoms at high temperatures.

As an indicator of correlation between potential energies of distant atoms, we have computed the mutual
information~\cite{CoverThomas2006}, which measures statistical dependence between two arbitrary random variables $x$ and $y$,
and is defined by
\begin{equation}
I = \int dx dy P(x, y|K)\ln \left[\frac{P(x, y|K)}{P(x|K)P(y|K)}\right],
\label{eq_mutual}
\end{equation}
where $P(x, y|K)$ is the joint probability distribution of $x$ and $y$, and $P(x|K)$, $P(y|K)$ are the marginal probability distributions.
The mutual information $I$ is non-negative, and strictly zero only for statistically independent variables, as in this case
$P(x, y|K)=P(x|K)P(y|K)$.

\begin{figure}[h!]
\begin{center}
\includegraphics[width=0.95\columnwidth]{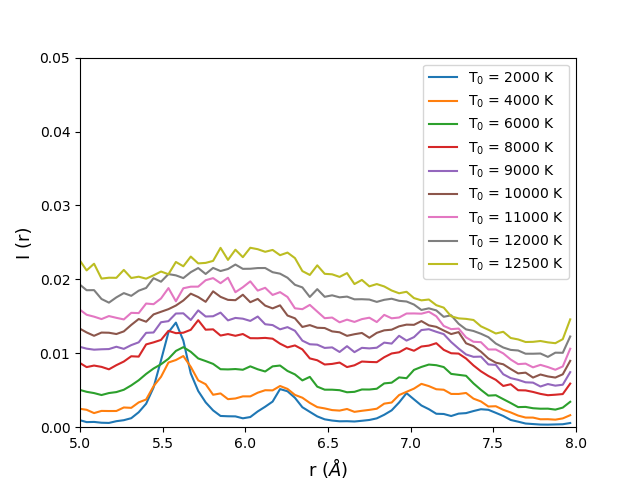}
\caption{Mutual information $I(r)$, as given in Eq. \ref{eq_mutual}, as a function of the interatomic distance $r$. At long-range the correlation measured by $I(r)$ increases as the solid becomes metastable.}
\label{fig_Mutinf}
\end{center}
\end{figure}

\begin{figure}[h!]
\begin{center}
\includegraphics[width=0.95\columnwidth]{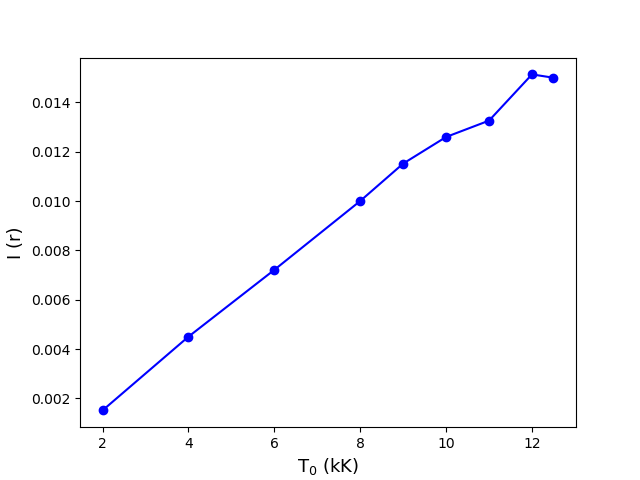}
\caption{Mutual information $I(r; T_0)$ as a function of the initial temperature $T_0$ at a fixed distance $r$ = 7.25 \AA. The values of $I(r; T_0)$ 
increase proportionally with the initial temperature, presenting a slight drop close to 12500 K.}
\label{fig_T0I}
\end{center}
\end{figure}

Fig. \ref{fig_Mutinf} shows the correlation, as measured by the mutual information diagnostic, between potential energies
$\phi_1$ and $\phi_2$ of a pair of atoms as a function of their distance $r$. It can be seen that for low temperatures,
around $T_0=$2000 K, the system presents only correlations for short distances, where the peaks are simply indicative of the
direct interaction of the atoms with their immediate neighbor shells. However, at larger distances this correlation decays
quickly, being almost zero beyond $r$ = 7.75 \AA. As the temperature is increased, the system gains energy and becomes
metastable once it crosses $T_m$, with a clear increase in correlation. In Fig. \ref{fig_T0I} the mutual information $I(r;
T_0)$ is shown  as a function of the initial temperature $T_0$ at a fixed distance of $r=$7.25 \AA. Interestingly, we can see
that the mutual information seems to exhibit a marked, almost linear dependence on $T_0$ which saturates close to $T_{LS}$.\\

At this point we have given a characterization of the superheated solid state in terms of some of its statistical properties. The emerging picture is one of increasing correlation between distant
atoms and reduced configurational degrees of freedom per atom as the system approaches $T_{LS}$ from below.
In addition, note that it is possible to establish similarities with the glassy state (supercooled liquid), a metastable state occurring at temperatures lower than the melting temperature. In both cases the system is located in points outside
the equilibrium phase, ``locked'' in a dynamical state that does not allow it to transition into the ``correct'' thermodynamical phase, and presenting nontrivial atomic kinetics.

Despite the fact that the case of superheated solid could seem opposite to the glassy state, apparently these transitions occur in a similar way, both involving cooperativity between their atoms, as the movement of one atom depends on the
movement of the others. In both cases there is the presence of spatial correlations near the phase transition, in our case near the melting point. This cooperative movement implies a reduction of the effective degrees of freedom, as shown in Figs. \ref{fig_dist3} and \ref{fig_T0I}. As the energy increases, correlations increase at large distances between the potential energies of the atoms, as shown in Fig. \ref{fig_T0I}, while in turn the number of effective degrees of freedom is reduced, as can be seen in Fig. \ref{fig_dist3}, because these correlations limit the amount of allowed movements of the atoms.

\section{Concluding remarks}
\label{concluding}

Using microcanonical molecular dynamics simulations we have characterized the statistics of single-atom potential energies in the metastable solid prior to melting,
according to a gamma distribution with energy-dependent parameters $k$ and $\theta$, which is consistent with correlated behavior. Furthermore, we have presented
strong evidence confirming the emergence of long-range spatial correlations of potential energy in the critical superheated solid. These correlations may
arise due to the cooperative dynamical processes occurring in the crystal after the creation of mobile vacancy-interstitial pairs.

\section*{Acknowledgements}

This work is supported by FONDECYT grant 1171127 and Anillo ACT-172101. Computations were performed using the supercomputing infrastructure of FENIX,
Materials Science Group, UNAB Physics Department (\url{http://www.matbio.cl/fenix}).

\appendix
\section*{APPENDIX: Microcanonical distributions of potential energy}

In the microcanonical ensemble, the probability density of microstates is given by

\begin{equation}
P(\bm{R}, \bm{P}|E) = \frac{\delta\Big(E-\sum_i\frac{p_i^2}{2m_i} - \Phi(\bm{R})\Big)}{\Omega(E)},
\end{equation}
with $\bm{R}=(\bm{r}_1, \ldots, \bm{r}_N)$, $\bm{P}=(\bm{p}_1, \ldots, \bm{p}_N)$ and where $\Omega(E)$ the density of states. By integration over $\bm{P}$ it can
be shown~\cite{Pearson1985,Ray1991,Davis2011a} that

\begin{equation}
P(\bm{r}_1, \ldots, \bm{r}_N|E) = \frac{1}{\eta(E)}\left[E-\Phi(\bm{r}_1, \ldots, \bm{r}_N)\right]_+^{\frac{3N}{2}-1},
\end{equation}
which in the limit $N \rightarrow \infty$ reduces to a canonical distribution,
\begin{equation}
P(\bm{r}_1, \ldots, \bm{r}_N|E) = \frac{1}{Z(\beta)}\exp(-\beta \Phi(\bm{r}_1, \ldots, \bm{r}_N)),
\label{eq_full_canon}
\end{equation}
with $\beta=3N/2E$. Therefore, the probability of observing a set of potential energies $\Phi_1=\phi_1$, $\Phi_2=\phi_2$,
\ldots, $\Phi_N=\phi_N$ at total energy $E$ is given by
\begin{align}
P(\phi_1, \ldots, \phi_N|E) = \frac{1}{Z(\beta)} \exp&\big(-\beta \sum_{i=1}^N \phi_i\big) \nonumber \\
                                                & \times\Omega_N(\phi_1, \ldots, \phi_N),
\label{eq_phi_vector}
\end{align}
where
\begin{displaymath}
\Omega_N(\phi_1, \ldots, \phi_N) = \int d\bm{r}_1\ldots d\bm{r}_N\prod_{i=1}^N\delta(\Phi_i-\phi_i)
\end{displaymath}
is the joint configurational density of states. For non-interacting atoms we have that $\Phi_i=\Phi_i(\bm{r}_i)$ and therefore
\begin{align}
\Omega_N(\phi_1, \ldots, \phi_N) & = \prod_{i=1}^N \left\{\int d\bm{r}\delta(\Phi_i(\bm{r})-\phi_i)\right\} \nonumber \\
                                 & = \prod_{i=1}^N \Omega_1(\phi_i),
\end{align}
hence $P(\phi_1, \ldots, \phi_N|E)$ factorizes and $\phi_i$ itself follows a canonical distribution
\begin{equation}
P(\phi_i | E) = \frac{1}{Z_1(\beta)}\exp(-\beta \phi_i)\Omega_1(\phi_i).
\end{equation}

Here $\Omega_1(\phi)=\int d\bm{r}\delta(\phi-\Phi_1(\bm{r}))$ plays the role of a ``density of states'' factor which is independent of the total energy $E$. On the other hand,
for interacting atoms the probability density $P(\phi_1|E)$ will be given by the marginalization of Eq. \ref{eq_phi_vector},
\begin{equation}
P(\phi_1|E) = \frac{1}{Z(\beta)}\exp(-\beta \phi_1)\omega_1(\phi_1; E)
\end{equation}
where
\begin{align}
\omega_1(\phi_1; E) = \int d\phi_2\ldots d\phi_N\exp& \Big(-\frac{3N}{2E}\sum_{i=2}^N \phi_i\Big) \nonumber \\
                                                    &\times \Omega_N(\phi_1, \ldots, \phi_N).
\end{align}

Now the ``density of states'' factor $\omega_1$ is energy-dependent. Similarly, for the joint distribution $P(\phi_1, \phi_2|E)$ we obtain by marginalization

\begin{equation}
P(\phi_1, \phi_2|E) = \frac{\exp\left(-\frac{3N}{2E}(\phi_1+\phi_2)\right)}{Z(\frac{3N}{2E})}\times \omega_2(\phi_1, \phi_2; E),
\label{eq_Pphi12}
\end{equation}
where $\omega_2(\phi_1, \phi_2; E)$ is defined by

\begin{align}
\omega_2(\phi_1, \phi_2; E) = \int d\phi_3\ldots & d\phi_N \exp\left(-\frac{3N}{2E}\sum_{i=3}^N \phi_i\right) \nonumber \\
      & \times \Omega_N(\phi_1, \ldots, \phi_N).
\end{align}


\end{document}